\documentclass[aps,prd,reprint]{revtex4-1}
\usepackage{savesym}
\usepackage{graphicx}
\expandafter\let\csname equation*\endcsname\relax
  \expandafter\let\csname endequation*\endcsname\relax 
\usepackage{subfig}
\usepackage{amsmath}
\usepackage{amssymb}
\usepackage{verbatim}
\usepackage[yyyymmdd,hhmmss]{datetime}
\usepackage{array}
\usepackage{times}
\usepackage[total={17.8cm,24.0cm},centering]{geometry} 
\usepackage{color}

\newcommand{\beq}{\begin{equation}}
\newcommand{\eeq}{\end{equation}}

\newcommand\ji {{J}}
\newcommand\ei {{\gamma}}
\newcommand\U {{U^r}}
\newcommand\dd {\partial}
\newcommand\D {{\cal D}}

\begin{document}

\title{ A complete characterisation of the orbital shapes of the non-circular Kerr geodesic solutions with circular orbit constants of motion  }

\author{Andrew Mummery}
\author{Steven Balbus}
\affiliation{Oxford University, Dept. of Physics, Beecroft Building, Keble Road, Oxford OX1 3 RH, United Kingdom}

\begin{abstract} 
We present analytical solutions describing a family of both inwardly and outwardly spiralling orbits in the Kerr spacetime.  The solutions are exact, and remarkable for their simplicity.    These orbits all have the angular momentum and energy of a circular orbit at some radius $r_c$, but are not restricted to remaining on that circular orbit, a property not possible in Newtonian gravity.   We demonstrate that there are five distinct orbital solutions which terminate at the black hole singularity, and three solutions which either escape to infinity or remain bound.  The different orbital solutions are characterised entirely by the black hole spin $a$ and the location of $r_c$.  Photon orbits spiralling into or out of their (unstable) circular orbit radii are also analysed.  These have properties similar to the hyperbolic class of massive particle orbits discussed herein. 
\end{abstract}
\maketitle
\section{ Introduction}


Despite the complexity of the Kerr spacetime, classes of exactly solvable timelike geodesics (orbital solutions) are known.    Examples include the familiar circular orbits,  radial plunges \cite{rev1}, ``zoom-whirl''  orbits, and homoclinic orbits, which separate bound and plunging orbital domains \cite{LP1}.    A useful compendium of such orbits may be found in \cite{LP2}.    The study of relativistic test-particle orbits characterised by the energy and angular momentum of a circular orbit, but which are not in fact circular, is by no means new.  As early as 1959, Darwin solved for some of these solutions in Schwarzschild geometry  \cite{Darwin}. The homoclinic orbits \cite{LP1} are another example of such ``pseudo-circular'' orbits, as are the inspirals from the innermost stable circular orbit of Kerr black holes derived recently \cite{MB}.  Still other studies of geodesic motion in black hole spacetimes are summarised in Chandrasekhar \cite{Chand}. While many such {\em particular} solutions are known, and completely general solutions of, e.g., the bound orbits of the Kerr spacetime are known in terms of elliptic integrals \cite{FH}, a complete characterisation of the entire family of the much simpler ``pseudo-circular'' orbital solutions is not present in the literature;  rather, incomplete sets of solutions are found in different papers. 

More than mathematical completeness is involved here.  Exact geodesic solutions  can be a powerful and extremely useful theoretical tool, and as such often surface in astrophysical studies. These include, for example, the study of black hole accretion flows \cite{NT, PT, Rey}, and the starting point for analysing the gravitational radiation from extreme mass ratio inspirals \cite{D, DH1, DH2, DH3, CH, LH, GHK}.  The latter is likely to be an important gravitational wave source for the LISA mission \cite{G}. Additions to the known family of simple orbital solutions are  therefore of general interest, as they offer insight into the internal dynamics of more complex situations, which are otherwise accessible only through numerical means. 

The purpose of the current study is to provide a systematic and detailed analysis of those orbits for which the effective potential of a Kerr black hole presents a double root and factors cleanly, a result of the underlying symmetries of the circular constants of motion.    This condition lends itself to integrability in terms of elementary functions.   As noted above, not all of the solutions presented in this paper are new, but the majority of them are.  In particular, we present a formal analysis of those orbits about extremal Kerr black holes $a = \pm M$ which do not appear to have been studied previously in the literature.  In particular, we demonstrate that eight distinct classes of these ``pseudo-circular'' orbital solutions exist, five of which plunge to the singularity, and three of which move to larger radii.  There, they may either remain bound or escape to infinity.  We highlight a number of interesting physical properties of these solutions.

The layout of this paper is as follows. In \S2  we introduce the general properties (stability, boundless, etc.) of circular test particle orbits in the Kerr metric. In  \S3 we demonstrate that associated with each of these circular orbits is a “pseudo-circular” orbit, which shares the same constants of motion (angular momentum and energy) as these circular orbits, but is not moving on a circular orbit. In  \S3 we derive the general properties of the radial velocity of these quasi-circular orbits.    In  \S4 we derive the solutions of the entire family of these orbits evolving in the Schwarzschild metric, before generalising to the Kerr metric in  \S5, and extremal Kerr black holes in \S6.  In \S7, we discuss how photon orbits fit into this general set of solutions, and show that they are analogous to a hyperbolic class of massive particle orbits. We conclude in \S8. 

\section{Preliminaries} 
We shall use geometric units in which the speed of light $c$ and the Newtonian gravitational constant $G$ are both set equal to unity.    In coordinates $x^\mu$, the invariant line element ${\rm d}\tau$ is given by
\beq
{\rm d}\tau^2 = - g_{\mu\nu} {\rm d}x^\mu {\rm d}x^\nu
\eeq
where $g_{\mu\nu}$ is the usual covariant metric tensor with spacetime indices $\mu, \nu$.   We shall use the standard Boyer-Lindquist coordinates $(t, r, \theta, \phi)$, where $t$ is time as measured at infinity and the other symbols have their usual quasi-spherical interpretation.    We shall work exclusively in the Kerr midplane $\theta =\pi/2$.
For black hole mass $M$, and angular momentum $a$ (both having dimensions of length in our choice of units), the non-vanishing $g_{\mu\nu}$ and $g^{\mu\nu}$ we require for our calculation may be summarised:
\begin{align}
g_{00} &= -1 +2M/r, \quad g^{00} = - {1 \over \Delta} \left(r^2 + a^2 + {2Ma^2 \over r }\right),  \nonumber \\
 g_{0\phi} &= g_{\phi0} = -2Ma/r, \quad g^{0\phi} = g^{\phi0} = -{2Ma} /{ r\Delta} , \nonumber \\
g_{\phi\phi} &= r^2+a^2 +2Ma^2/r, \quad g^{\phi\phi} = {1\over \Delta} \left(1 - {2M\over r}\right), \nonumber \\ 
g_{rr} &= r^2/\Delta, \quad \Delta \equiv r^2 - 2Mr +a^2. 
\end{align}

The 4-velocity vectors are denoted by $U^\mu={\rm d}x^\mu/{\rm d}\tau$.    For circular orbits, only $U^0$ and $U^\phi$ are present,    their $r$-dependence may be deduced from the two coupled algebraic equations:
\beq
-1 = g_{\mu\nu} U^\mu U^\nu,\quad 0= U^\mu U^\nu \partial_r g_{\mu\nu}
\eeq
where $\partial_r$ is the standard partial derivative with respect to $r$.   (The second equation is a consequence of the geodesic equation for ${\rm d}U_r/{\rm d}\tau$; see e.g., \cite{HEL}).    Solving these equations gives 
\beq
U^\phi = {M^{1/2}\over r^{3/2}\D}, \quad U^0 =  {1+ a M^{1/2} / r^{3/2} \over \D} ,
\eeq
where $\D^2=1 -3M/r +2aM^{1/2}/r^{3/2}$.

Circular values of angular momentum and energy constants are calculated in their usual manner ($U_\mu = g_{\mu\nu}U^\nu$) and are given by
\begin{align}
\ji \equiv U_\phi &= (Mr)^{1/2} {(1+a^2/r^2-2aM^{1/2}/r^{3/2}) \over \sqrt{1 -3M/r +2aM^{1/2}/r^{3/2}}}, \\
\ei \equiv -U_0 &=  { (1-2M/r +aM^{1/2}/r^{3/2})  \over \sqrt{ 1 -3M/r +2aM^{1/2}/r^{3/2}}}.
\end{align}

\section{Radial velocity of pseudo-circular orbits}

\subsection{The radial velocity equation}
The radial velocity of a massive particle is given by the solution of the geodesic equation $g_{\mu\nu}U^\mu U^\nu = -1$, or explicitly 
\begin{multline}\label{full}
(\U)^2 + {\ji \over r^2} \left( {2Ma\ei \over r} + \left(1 - {2M \over r}\right) \ji \right) \\ - {\ei \over r^2} \left(\left(r^2 + a^2 + {2Ma^2 \over r}\right) \ei - {2Ma\ji \over r}\right)\\ + 1 + {a^2 \over r^2} - {2M \over r} = 0 ,
\end{multline}
where we have expressed all azimuthal $U^\phi$ and temporal $U^0$ 4-velocity components in terms of their (conserved) 4-momenta counterparts $J$ and $\gamma$ \cite{MB}. 
Note that equation \ref{full} is of the form $(\U)^2 + V_{\rm eff}(r) = 0$, which defines an effective potential $V_{\rm eff}$, cubic in $1/r$, which may generally be factored:  
\beq
 V_{\rm eff}(r) = - V_0 \left({r_1\over r} - 1\right)\left({r_2\over r}- 1\right)\left({r_3\over r}-1\right),
\eeq 
where $r_1, r_2$ and $r_3$ are the general (possibly complex) roots of $\U$. 
For a circular orbit at radius $r_c$,  both $V_{\rm eff}(r_c) = 0$ and $\partial_r V_{\rm eff}(r_c) = 0$, meaning $r_c$ must be a double root of $V_{\rm eff}$, i.e.,  $r_1=r_2=r_c$.  The normalisation constant $V_0$ may be found by going back to equation (\ref{full}), and evaluating $r^3(\U)^2$ in the limit $r\rightarrow 0$. Taking this limit we find
\beq
V_0 =  {2M \over r_1 r_2 r_3} (\ji - a\ei)^2 .
\eeq   
On the other hand, we may consider the $r \rightarrow \infty$ limit, which leads to
\beq
V_0 = 1 - \ei^2 .
\eeq 
Taking $\ei$ and $\ji$ to be the energy and angular momentum of a circular orbit at $r_c$, and writing $r_1 = r_2 = r_c$, we can constrain the other root $r_3$ with these two expressions
\beq\label{rstar}
r_3 = {2M \over r_c^2} {(\ji - a\ei)^2 \over 1 - \ei^2},
\eeq  
which gives the radial velocity equation
\beq\label{urgen}
U^r = - \sqrt{1 - \ei^2} \left({r_c \over r} - 1\right) \sqrt{{r_3 \over r} - 1} . 
\eeq
For any given black hole spin and mass, $\gamma$, $J$ and $r_3$ are fully determined by the choice of $r_c$.  As $r_c$  descends through smaller radii, the behaviour of pseudo-circular orbits qualitatively changes as $r_c$ passes through a series of characteristic radii, which we discuss below. 

\subsection{Characteristic radii} 
The largest characteristic radius is the innermost stable circular orbit, or ISCO, denoted $r_I$.   Circular orbits with $r_c > r_I$ are stable and approach their Newtonian behaviour at large radii; orbits with $r_c< r_I$ are dynamically unstable. The ISCO itself corresponds to a radius of marginal stability. 

The second key radius is the innermost bound circular orbit, $r_{\rm ib}$, or IBCO.   Circular orbits with $r_{\rm ib}< r_c <r_I$ remain bound though they are unstable.  Physically, this instability results in a class of pseudo-circular orbits that describe a particle infinitesimally perturbed outwards from $r_c$, then traveling out to a finite radius ($r_3$) before returning to $r_c$ once again in the limit $t\to\infty$.  

The final key circular orbit radius is the photon radius $r_p$, lying interior to $r_{\rm ib}$.  The photon radius corresponds to the unique radius at which photons are able to undergo (unstable) circular motion.     Massive particle circular orbits with $r_p < r_c< r_{\rm ib}$ are possible, but are both unstable and ultimately unbound. The region between $r_p$ and the outer event horizon $r_+$ does not host any circular orbit solutions, for either massive particles or for photons,  as this would require superluminal velocities.

\begin{figure}
\includegraphics[width=\linewidth]{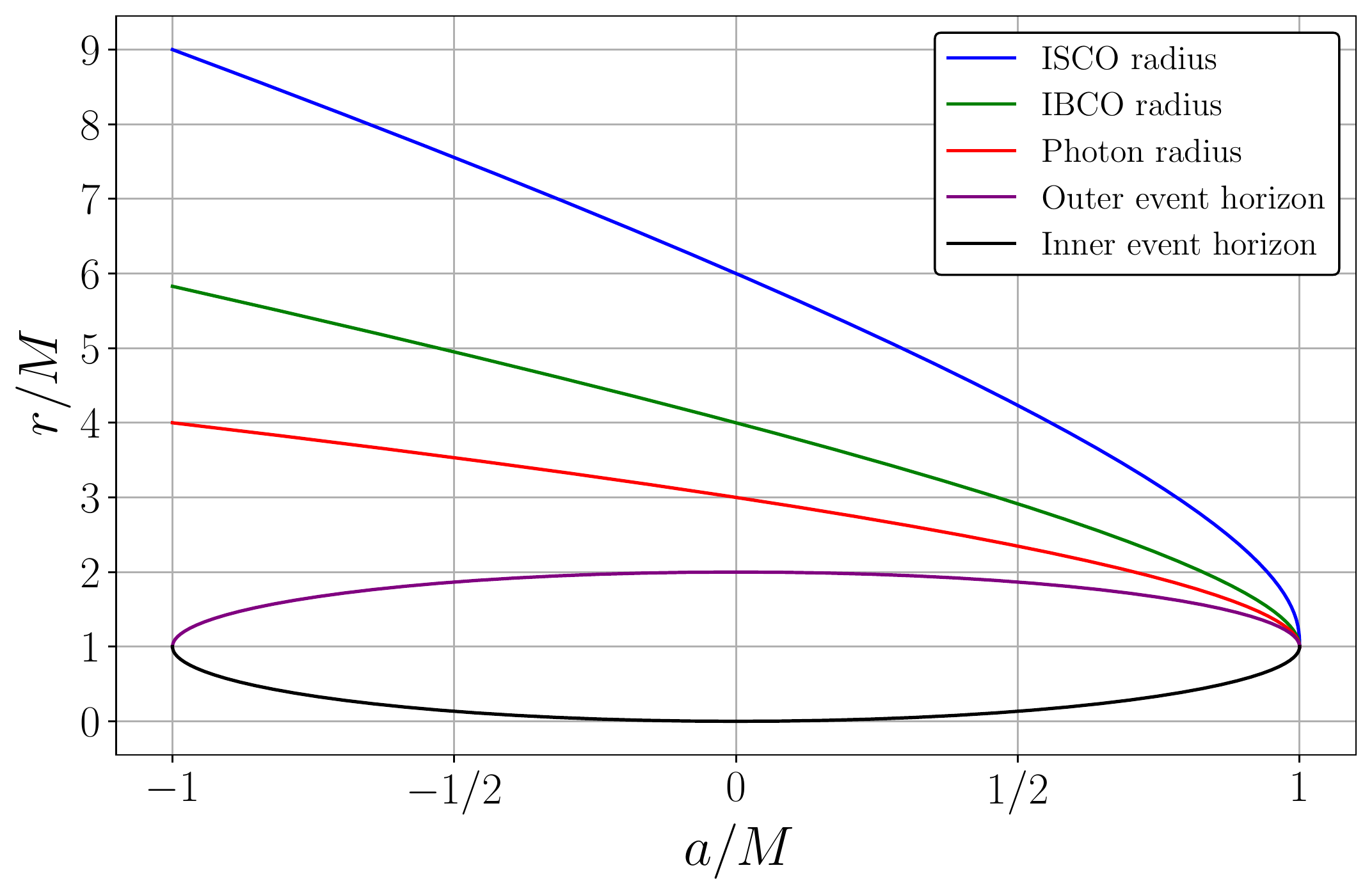}
\caption{The key radii of the Kerr spacetime, plotted as a function of black hole spin.  }
\label{kerr_radii}
\end{figure}

When normalised to $M$, each of these special radii depends only upon the Kerr black hole's angular momentum parameter $a$ (e.g., \cite{HEL}) via $a_\star = a/M$.  The ISCO radius is given explicitly  by \cite{Bard}
\beq
{r_I \over M} = 3 + Z_2 - {a_\star \over |a_\star|}  \sqrt{(3-Z_1)(3 + Z_1 + 2  Z_2)} ,
\eeq
where
\beq
Z_1 = 1 + \left(1-a_\star^2\right)^{1/3}  \left[ \left(1+a_\star \right)^{1/3} + \left(1-a_\star \right)^{1/3} \right] ,
\eeq
and
\beq
Z_2 = \sqrt{3a_\star^2 + Z_1^2} .
\eeq
The IBCO radius is \cite{Chand}
\beq
{ r_{\rm ib} \over M} = \left(1 + \sqrt{1 - a_\star} \right)^2 .
\eeq
The photon radius is given by \cite{HEL}
\beq
{r_p \over M} = 4 \cos^{2}\left[{1\over3}\cos^{-1}\left(-{a \over M}\right)\right] ,
\eeq
and finally, the inner ($r_-$) and outer ($r_+$) event horizons of a Kerr black hole are located at \cite{HEL}
\beq
{r_{\pm} \over M} = 1 \pm \sqrt{1 - a_\star^2} .
\eeq
The values of these five important radii are plotted as a function of black hole spin in Fig. \ref{kerr_radii}.  In the formal $a_\star \to 1$ limit, each of these radii converge on $r = M$.   However, this is in actual fact a Boyer-Lindquist coordinate effect.   They key radii remain separated by a finite {\it proper} radial distance. A free-falling observer would, just as for all other values of the black hole spin parameter, first pass an ISCO, then the IBCO, then the photon orbit, and finally the event horizon in this $a = M$ limit.    

\subsection{The relationship between  $r_3$ and $r_c$.}

Figure 1 shows that the ordering $r_I>r_{\rm ib} > r_p$ is always obeyed.   The nature of a circular orbit will then depend upon the magnitude of $r_c$ relative to these key quantities, and is best understood by using $r_c$ and $r_3$ (a direct function of $r_c$) in tandem.


Consider first the large $r_c$ limit of the $r_3$ equation. We then have $\ji \rightarrow \sqrt{Mr_c}$ and $1 - \ei^2 \rightarrow M/r_c$ following equations 5 and 6.   Equation \ref{rstar} then demonstrates that $r_3$ approaches $2M$, the Schwarzschild event horizon.  The same limit holds for Kerr black holes as well, and so lies outside $r_+$ for all values of $a_\star$.    

\begin{figure}
\includegraphics[width=\linewidth]{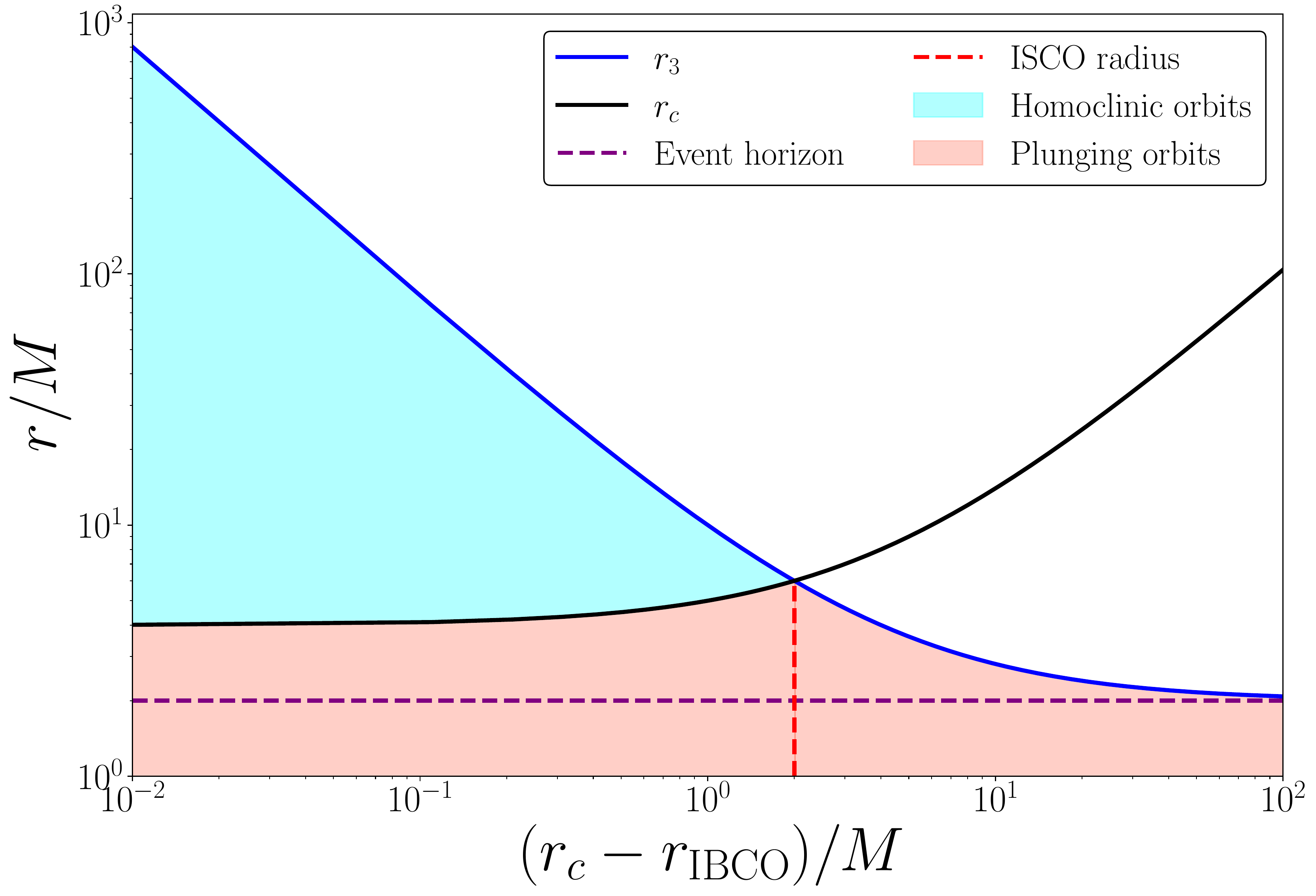}
\caption{The properties of the  third root $r_3$ of the Schwarzschild ($a=0$) velocity potential, plotted against the difference of the circular radius $r_c$ from the IBCO $r_{\rm ib} = 4M$. Different regions of radial space available to different classes of pseudo-circular orbits are indicated by the shaded regions.  See text for further details.  }
\label{schwarz_rstar}
\end{figure}

\begin{figure}
\includegraphics[width=\linewidth]{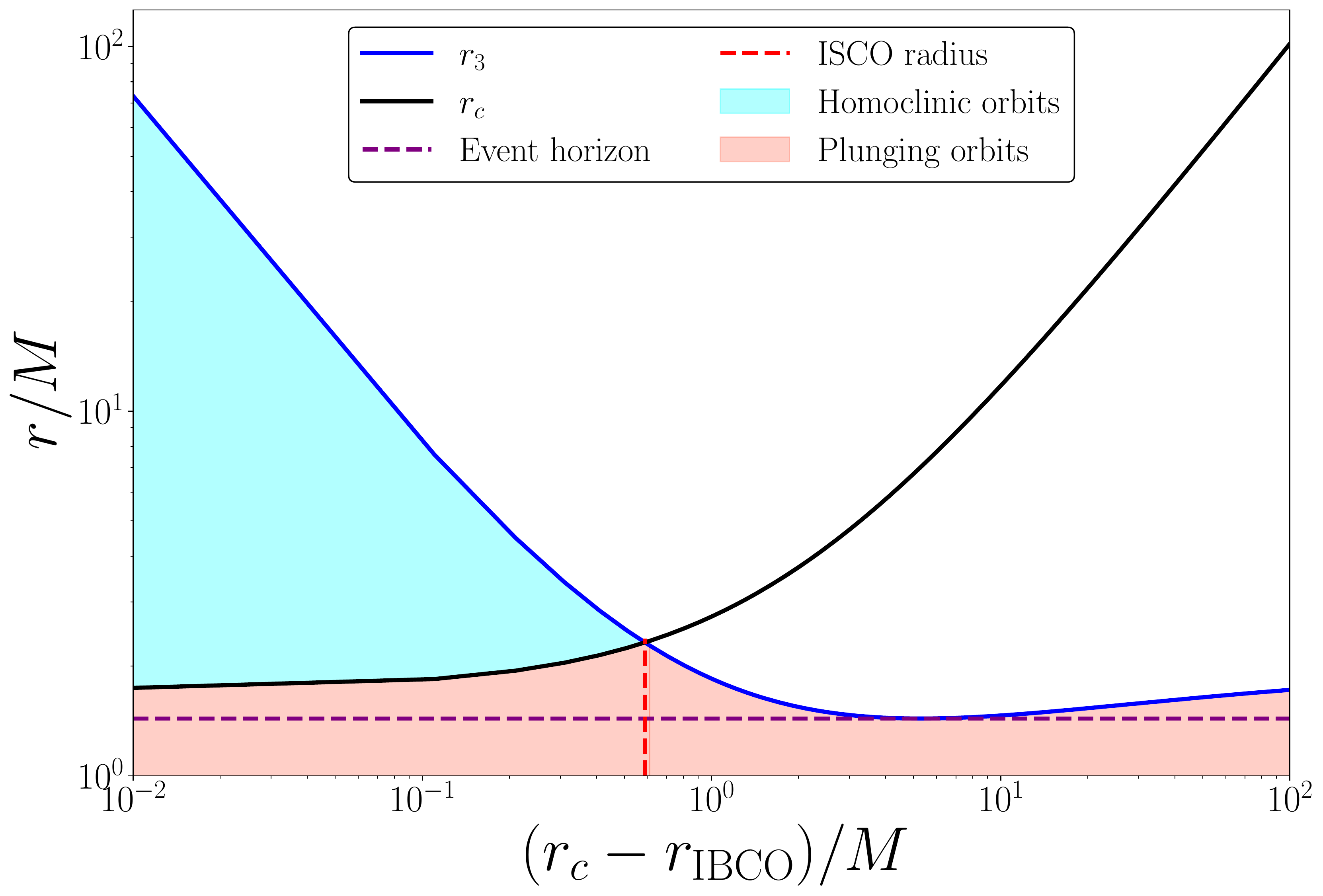}
\caption{As in Fig. \ref{schwarz_rstar}, except for Kerr spin parameter $a = 0.9M$. Note that $r_3 \geq r_+$ for all values of both $r_c$ and $a$.  }
\label{kerr_rstar}
\end{figure}

Figs. \ref{schwarz_rstar} and \ref{kerr_rstar} are plots in the $(r_c-r_{\rm ib})/M$,  $r/M$ plane, for a Schwarzschild and $a=0.9$ Kerr black hole respectively.   (We restrict the plots to the $r_c > r_{\rm ib}$ domain.)   Both $r_c$ and $r_3$ appear as labelled curves, which by definition must intersect at $r_c=r_I$.      It can be seen that for $r_c > r_I$, the root $r_3$ is  always located between $r_+$ and $r_I$.  In particular,  as $r_c$ approaches $r_I$ from above, $r_3$ approaches $r_I$ from below.   At the intersection, the right side of equation 12 for the radial velocity takes the form of a  square root of a perfect cubic.   This is of some astrophysical interest \cite{MB}.  
As $r_c$ approaches $r_{\rm ib}$, the $r_3$ root increases monotonically, diverging  as $r_c \rightarrow r_{\rm ib}$.  

The shaded zones of these plots denote regions accessible to a particle in pseudo circular orbit for a given $r_c$ (or the associated $r_3$).    The red region indicates the domain of plunging orbits; the blue, homoclinic orbits. (Homoclinic orbits will be discussed in more detail below.)   In the white zone, only $r=r_c$ is allowed, in conformance with Newtonian sensibilities!   For $r_c = r_{\rm ib}$, the particle may access all radii whilst retaining its constant angular momentum and energy, either through an escape to infinity with infinitesimal final velocity, or a plunge into the singularity.   Finally, we find $r_3 < 0$ when $r_c$ is located between the IBCO and photon radius,  $r_{p }< r_c < r_{\rm ib}$.  (Not shown.)   In this regime, all radii are similarly accessible, the difference being that particles escaping to infinity do so with finite velocity.  

\section{Schwarzschild solutions}\label{schwarz}
The Schwarzschild metric is particularly simple as, unlike the finite $a$ Kerr metric, the $\phi$ coordinate is well-behaved at the single event horizon.   The general solution required for characterising the various Schwarzschild orbits starts with the simple expression for the $\phi$ component of the test particles 4-velocity 
\beq
{{\rm d} \phi \over {\rm d} \tau} = U^\phi = {\ji  \over r^2} .
\eeq
The shape of a particle's orbit is then given by the solution of the following ordinary differential equation
\beq
{{\rm d} \phi \over {\rm d}r } = {U^\phi \over \U} =-  {\ji  \over  \sqrt{1 - \ei^2}} {1 \over r^2  \left({r_c / r} - 1\right) \sqrt{{r_3 / r} - 1} } ,
\eeq
or explicitly 
\beq\label{phi_int_1}
 {\sqrt{1 - \ei^2} \over \ji } \phi = - \int  {{\rm d} r \over r^2}  {1 \over   \left({r_c / r} - 1\right) \sqrt{{r_3 / r} - 1} } .
\eeq
For three of the five radial regimes we may solve the integrals by defining $\tilde u = r_3 /r$, leaving 
\beq
 {\sqrt{1 - \ei^2} \over  \ji} \phi = {1 \over r_3} \int     {1 \over   \left({r_c \tilde u / r_3  } - 1\right) \sqrt{{ \tilde u } - 1} } \, {\rm d} \tilde u .
\eeq
Finally, with $y = \sqrt{\tilde u - 1}$ we have
\beq
 {\sqrt{1 - \ei^2} \over  \ji} \phi = {2 \over r_c} \int     {1 \over   \left(  1 - r_3 / r_c + y^2  \right)  } \, {\rm d} y ,
\eeq
note  $r = r_3 / (1 + y^2)$. The exact solution of this final integral depends on the value of the ratio $r_3/r_c$. The integral
\beq
I(y) =  \int     {1 \over   \left(  1 - r_3 / r_c + y^2  \right)  } \, {\rm d} y ,
\eeq
has solution
\begin{align}
I(y) = 
\begin{cases}
 \tan^{-1}\left(y / \sqrt{1 - r_3 / r_c}\right) /  \sqrt{1 - r_3 / r_c},\quad  r_3 < r_c, \\\\
 -1/y ,\quad   r_3 = r_c, \\\\
  -\tanh^{-1}\left(y / \sqrt{ r_3 / r_c - 1}\right) /  \sqrt{ r_3 / r_c - 1},\quad  r_3 > r_c.  
\end{cases}
\end{align}
For the final radial regime $r_p < r < r_{\rm ib}$, we have $\ei > 1$ and $r_3 < 0$ and the integral (eq. \ref{phi_int_1}) should be re-written before the substitution $\tilde u = -r_3/r = |r_3|/r$ is made. Writing all terms as a manifestly positive,
\beq
 {\sqrt{\ei^2 - 1} \over \ji } \phi = - \int  {{\rm d} r \over r^2}  {1 \over   \left({r_c / r} - 1\right) \sqrt{{|r_3| / r} + 1} }.
\eeq
We now follow a similar approach to the previous case with $\tilde u = -r_3/r = |r_3|/r$, $y = \sqrt{1 + \tilde u}$, and $r = |r_3| / (y^2 - 1) $.
We then find 
\beq\label{phi_int_2}
 {\sqrt{\ei^2 - 1} \over \ji } \phi =  \int   {1 \over   \left({1 + |r_3| / r_c} - y^2\right)  } {\rm d} y .
\eeq
Note 
\beq
\int   {1 \over   \left({1 + |r_3| / r_c} - y^2\right)  } {\rm d} y = {\tanh^{-1}\left(y / \sqrt{ |r_3| / r_c + 1}\right) \over  \sqrt{ |r_3| / r_c + 1}}.
\eeq
Note that $\tanh^{-1}(X)$ and $\tanh^{-1}(1/X)$ have exactly the same derivative, so that the proper argument of this function must be chosen on the basis of whether $X$ or its reciprocal is less than unity.   
This flexibility embodies the difference between the plunging and out-spiralling orbits. 
We now discuss each of the eight different solutions of these integrals in turn. 

\subsection{$r_c > r_{I}$ plunging orbits}
\begin{figure}
\includegraphics[width=\linewidth]{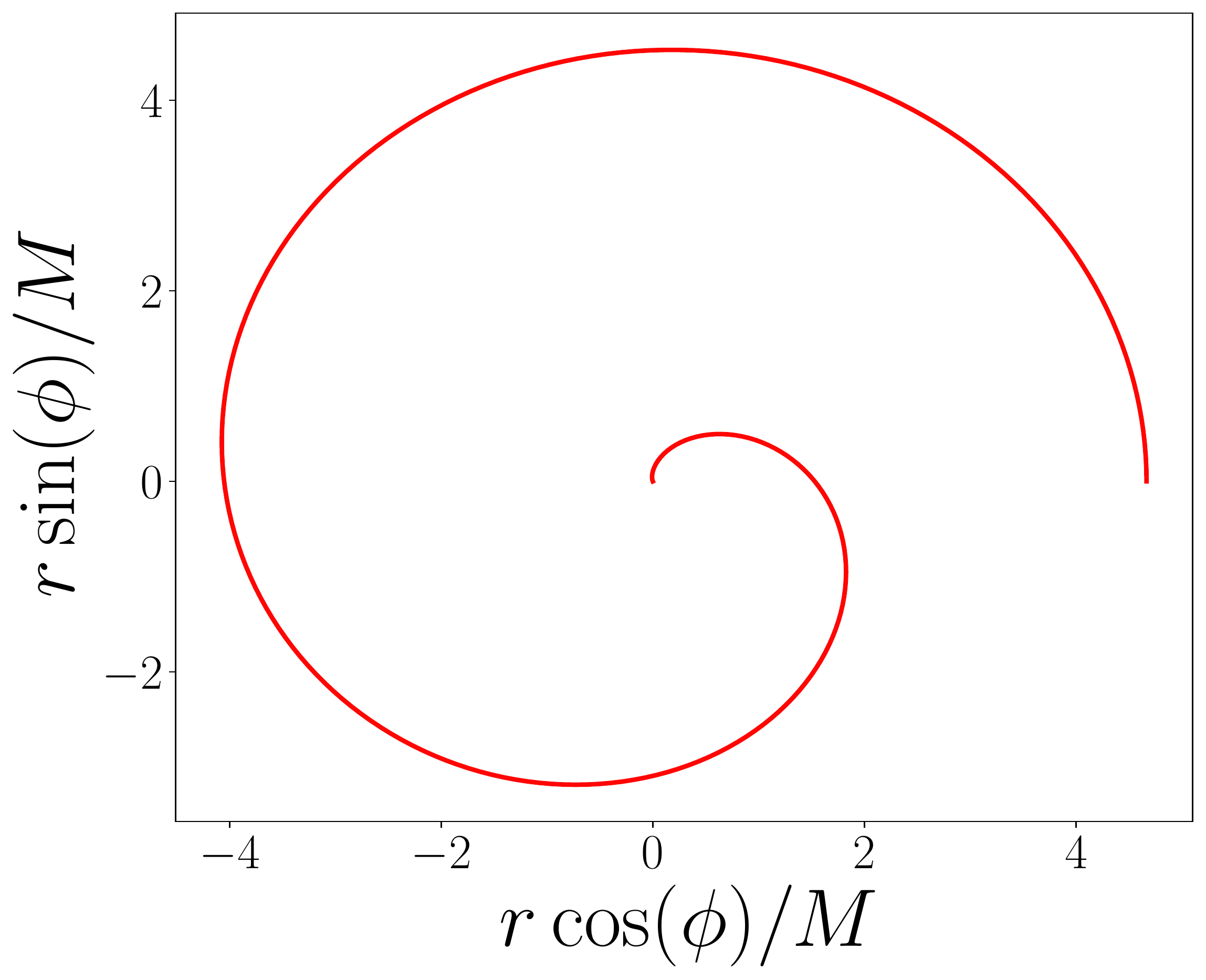}
\caption{A Schwarzschild plunging orbit, with $r_c = 7M$. }
\label{SP}
\end{figure}

For $r_c > r_I$, we have $r_3 < r_c$.  We may then solve explicitly for $r(\phi)$ from the first line of equation 25, 
\beq
r(\phi) = {r_3 \over 1 + (1 - r_3 / r_c) \tan^2 (\phi/\phi_\star)} ,
\eeq
where 
\beq
\phi_\star = {2 \ji \over \sqrt{r_c(r_c - r_3) (1 - \ei^2)}} ,
\eeq
with the convention that the orbit starts at $\phi = 0$.   Unlike other Schwarzschild solutions, the $r_c >r_I$ orbits involve only a finite rotation angle $\Delta\phi$ before reach the singularity $r=0$ (Fig. \ref{SP}): 
\beq
\Delta \phi = {\pi \ji \over \sqrt{r_c(r_c - r_3) (1 - \ei^2)}} .
\eeq
Note the divergence as $r_3 \rightarrow r_c$ (i.e., $r_c \to r_I$).   

As $r_c \rightarrow \infty$, we have
\beq
\Delta\phi \rightarrow \pi \left(1 - {2M \over r_c}\right)^{-1/2} > \pi,
\eeq
i.e. all orbits complete at least one half-rotation of the singularity, independent of $r_c$.
Interestingly, in this limit the plunging particle has a formally divergent specific angular momentum $J \rightarrow \sqrt{Mr_c} \rightarrow \infty$.   Having an arbitrarily large amount of angular momentum and starting outside the event horizon is not sufficient for preventing the particle from eventually being swallowed by the singularity!


\subsection{$r_c = r_{I}$ spiral orbits}
The ISCO radius $r_c = r_I$ is the special location at which $r_3 =r_c = 6M$, and the effective potential is a perfect cube. This remarkably simple solution was discussed in detail in \cite{MB}, and seems to have been first written explicitly in \cite{Chand}.  It has the following form:
\beq
r(\phi) = {6M \over 1 + 12/\phi^2} ,
\eeq
with the convention that $\phi$ increases from $-\infty$ at $r=6M$ to $0$ at the singularity $r=0$. 

\subsection{$r_{\rm ib} < r_c < r_{I}$ homoclinic and plunging orbits }
For $r_{\rm ib} < r_c < r_I$,  two possible pseudo-circular orbits exist.  These correspond to orbits which either spiral outwards from $r_c$ towards $r=r_3$, or inwards from $r=r_c$ towards $r=0$. 

\subsubsection{$r > r_c$: homoclinic orbits}
The homoclinic orbits where first discussed in full detail by \cite{LP1}.
However, the conceptual history of homoclinic orbits goes back much further in time.  Darwin \cite{Darwin} first wrote down an explicit Schwarzschild homoclinic orbit in 1959.  (Interestingly, Darwin writes that he was motivated by his experience with the relativistic Bohr model for hydrogen.)    While these solutions are therefore not new, we shall include a brief discussion for completeness.  

For $r_{\rm ib} < r_c < r_I$, the third root of the radial velocity $r_3$ is greater than $r_c$. Thus, orbits exist where the particle spirals outwards from $r=r_c$ towards $r=r_3$.  The following solution for $r(\phi)$ describes the test particles motion:
\beq
r(\phi) = {r_3 \over 1 + ( r_3 / r_c - 1) \tanh^2 (\phi/\phi_\star)} ,
\eeq
where 
\beq\label{phistar}
\phi_\star = {2 \ji \over \sqrt{r_c(r_3 - r_c) (1 - \ei^2)}} . 
\eeq
A homoclinic orbit begins with $\phi = -\infty$, at $r = r_c$. As the azimuthal coordinate increase the radial coordinate grows until  $\phi = 0, r = r_3$, before the orbit repeats itself, with the radius returning to $r=r_c$ as $\phi \rightarrow + \infty$.

\begin{figure}
\includegraphics[width=1.15\linewidth]{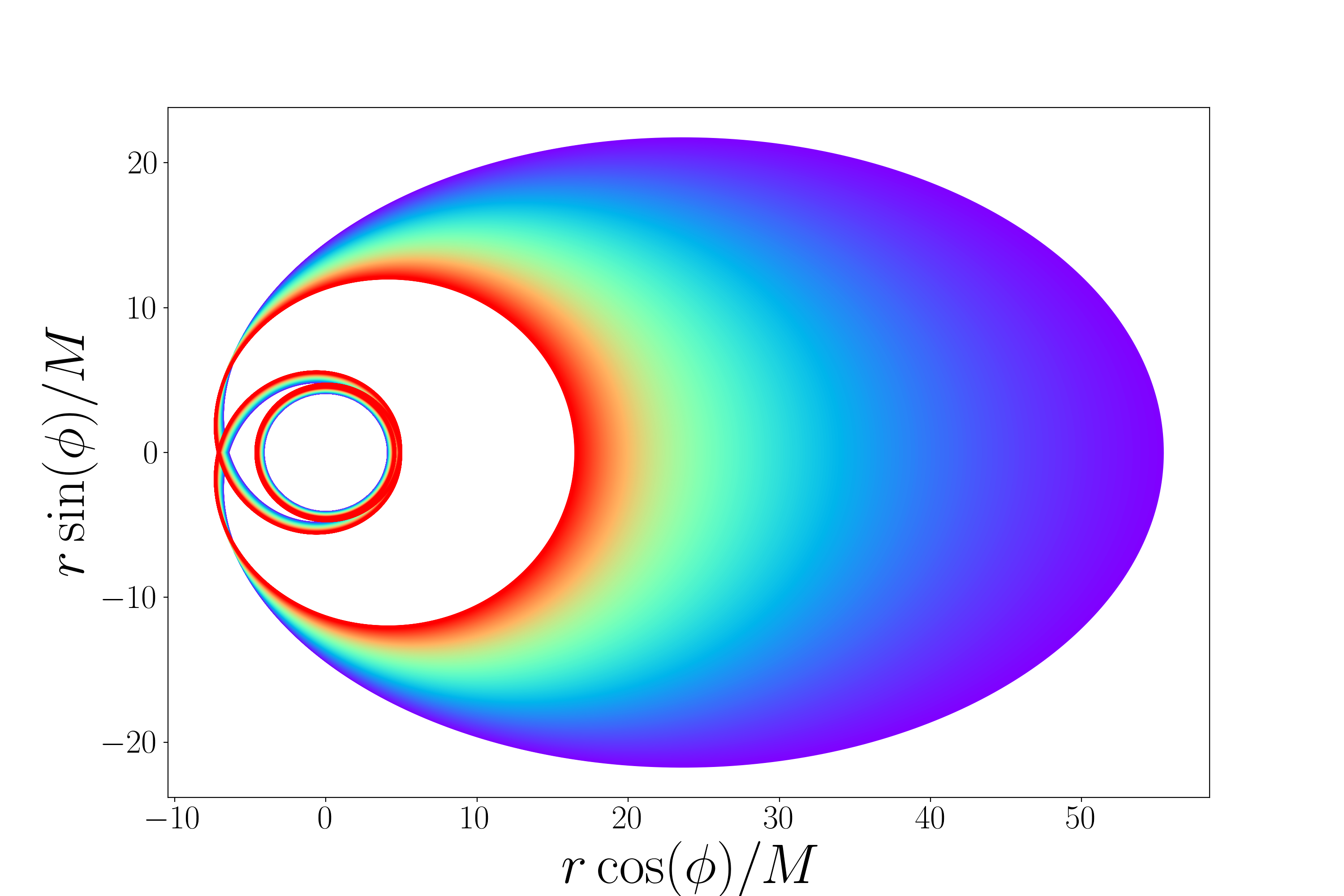}
\caption{A series of many Schwarzschild homoclinic orbits, with circular orbit radii ranging from $r_c = 4.15M$ (purple) to $r_c = 4.45M$ (red). }
\end{figure}

\subsubsection{$r < r_c$: plunging orbits}
The complementary test particle orbit which spirals inwards from $r=r_c$ to the singularity is described by the following function: 
\beq
r(\phi) = {r_3 \over 1 + ( r_3 / r_c - 1) \coth^2 (\phi/\phi_\star)} ,
\eeq
where $\phi_\star$ is given once again by eq. once again by eq. \ref{phistar}.  
This orbit is in many ways similar to the inspiralling trajectory starting from the ISCO. The particle undergoes a formally infinite number of rotations about the unstable circular orbit $r_c$ (the $\phi \to -\infty$ limit), before transitioning to a plunge which in effect ``begins'' at an azimuthal angle $\phi \sim - \phi_\star$ and terminates at $\phi = 0$. 

\subsection{$r_c = r_{\rm ib}$ parabolic orbits}
\begin{figure}
\includegraphics[width=1.\linewidth]{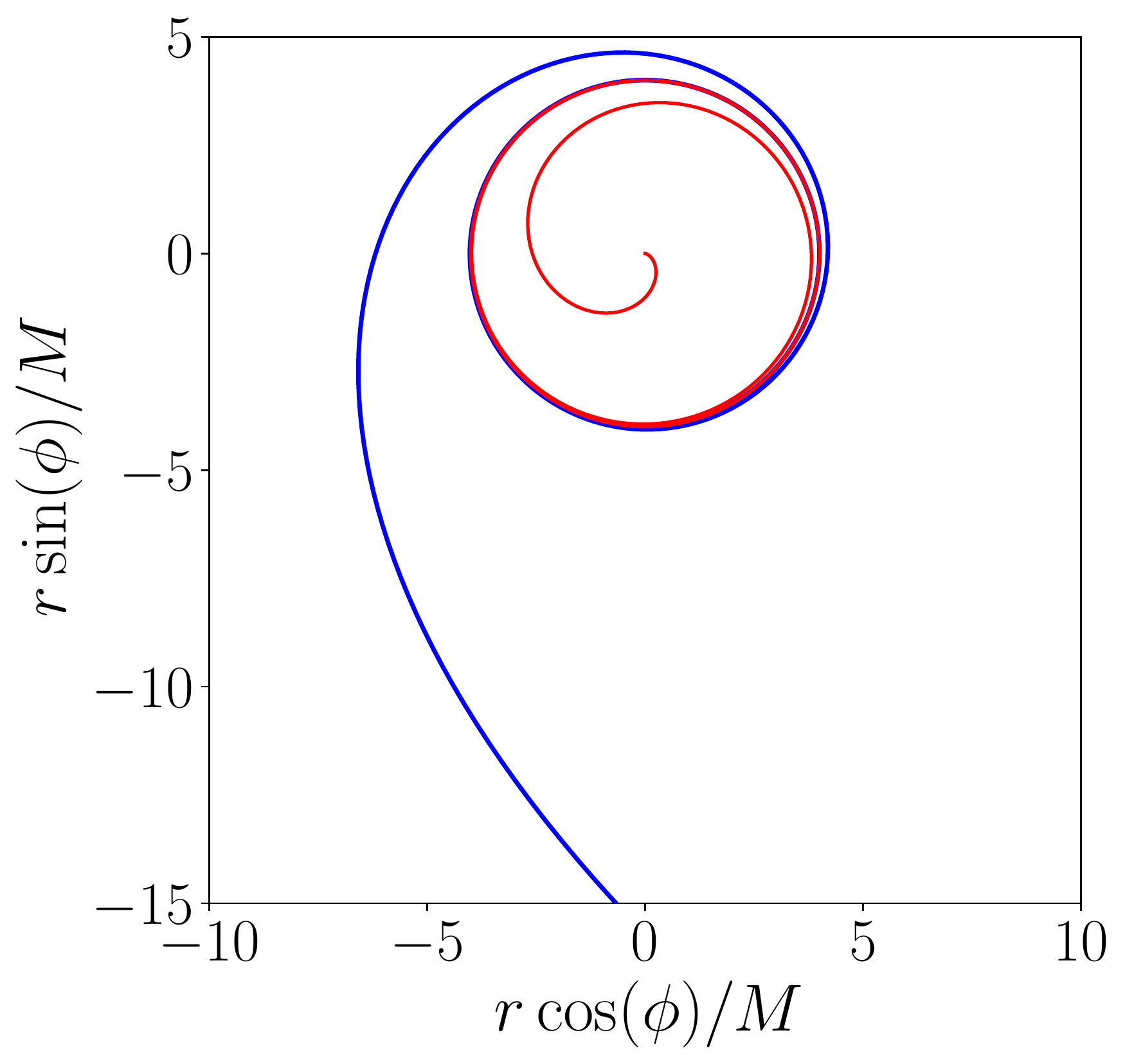}
\caption{The Schwarzschild parabolic orbits, spiralling outwards (blue) or inwards (red) from the IBCO ($r_{\rm ib}=4M$). }
\label{schwarz_parabola}
\end{figure}

The IBCO is defined by the circular orbit radius at which  $\ei(r_{\rm ib}) = 1$, allowing an escape to infinity with vanishing velocity, and $r_3 \rightarrow \infty$.  In this limit, the radial component of the test particle 4-velocity is quite well-behaved, and simplifies to 
\beq
U^r_{\rm ib} = -{ J \over r_{\rm ib}}    \sqrt{2M \over r} \left({r_{\rm ib} \over r} - 1\right) .
\eeq
For a Schwarzschild black hole, from equations 5 and 6 we find $J_{\rm ib} =  r_{\rm ib} = 4M$, and $U^\phi = 4M/r^2$.   Thus 
\beq
{{\rm d}\phi \over {\rm d}r} = {U^\phi \over U^r} = {2\sqrt{2M} \over r^{3/2}}{1 \over \left({4M/ r} - 1\right)} .
\eeq 
The two solutions of this differential equation (corresponding to inspiralling and outspiralling orbits) are 
\begin{equation}
r(\phi) = 4M \tanh^{2}\left( {\phi\over2\sqrt{2}}\right), \quad (r < r_{\rm ib} \,\, {\rm inspiral}) ,  
\end{equation}
and 
\begin{equation}
r(\phi) = 4M \coth^{2}\left( {\phi\over2\sqrt{2}}\right), \quad (r > r_{\rm ib} \,\, {\rm outspiral}) .   
\end{equation}
These are displayed in Fig. \ref{schwarz_parabola}. Note that by running the out-spiralling orbit in a time reversed fashion, we see a particle in-falling  from rest at infinity, and ending on a circular orbit. 

\subsection{$r_{p} < r_c < r_{\rm ib}$ hyperbolic orbits}
\begin{figure}
\includegraphics[width=1.\linewidth]{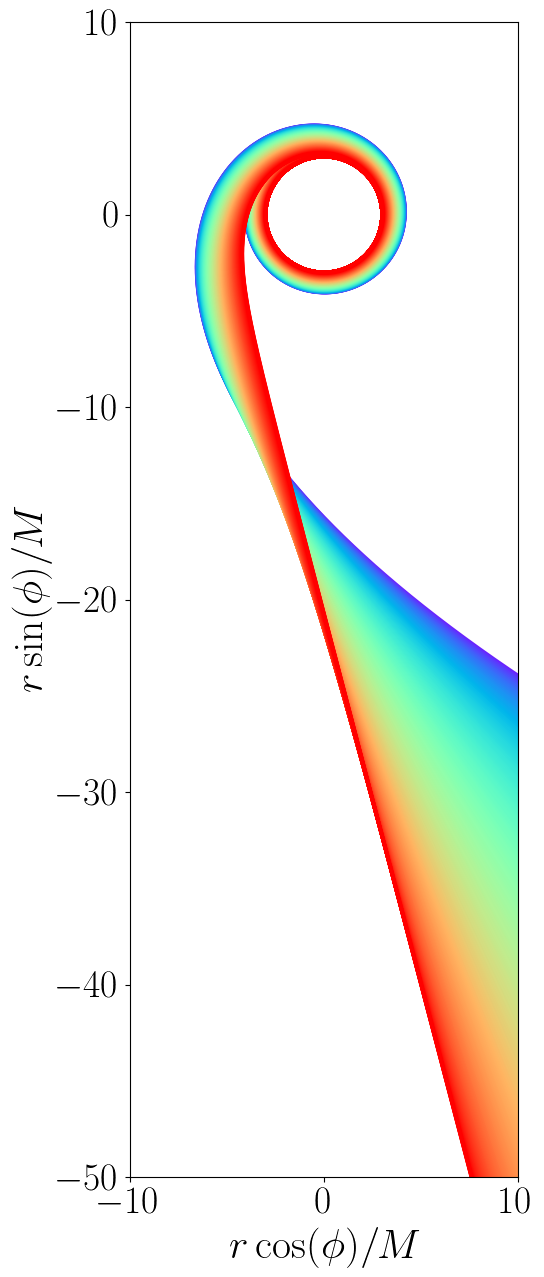}
\caption{A series of many Schwarzschild hyperbolic orbits, with circular orbit radii ranging from $r_c = r_{\rm ib} - 0.01M$ (purple) to $r_c = r_p + 0.01M$ (red). }
\label{schwarz_hyperbol}
\end{figure}

 For $r_{p} < r_c < r_{\rm ib}$ we have $\gamma > 1, r_3 < 0$, and a test particle can escape to infinity with non-zero radial velocity.  The shape of the particles orbit is the following 
 
 \beq
 r(\phi) = {|r_3| \over ( 1 + |r_3| / r_c) \tanh^2(\phi/\phi_\star) - 1}, \quad ({\rm outspiral})
 \eeq
 where 
\beq
\phi_\star = {2 \ji \over \sqrt{r_c(|r_3| + r_c) ( \ei^2 - 1 )}} . 
\eeq
 A test particle undergoing this orbit escapes to infinity at a finite angle (the orbit begins with $\phi \to -\infty$) 
 \beq
 \phi_\infty = -{2 \ji \over \sqrt{r_c(|r_3| + r_c) ( \ei^2 - 1 )}} \tanh^{-1} \left({1 \over \sqrt{1 + |r_3|/r_c}}\right) .
 \eeq
This escape angle is a measure of the finite radial velocity with which these orbits approach infinity.  These trajectories are plotted in Fig. \ref{schwarz_hyperbol}. The decrease in the escape angle as $r_c$ approaches the photon radius $3M$ is clearly visible. 
 Alternatively, the test particle may spiral inwards from $r=r_c$ to the singularity at $r=0$ according to the following profile
 \beq
 r(\phi) = {|r_3| \over ( 1 + |r_3| / r_c) \coth^2(\phi/\phi_\star) - 1}, \quad ({\rm inspiral}) . 
 \eeq
with the convention where $\phi$ progresses from $-\infty$ to $0$ while $r$ progresses from $r_c$ to $0$.

\section{Kerr solutions}
The shape of spiral in the Kerr metric for a finite mass particle differs from the equivalent Schwarzschild orbits in two key ways: the first is that frame dragging forces the particle to co-rotate with the black hole within the ergosphere, which in the equatorial plane lies at the location $r_E = 2M$. The second is that the $\phi$ diverges, a purely bad-coordinate effect, at the two event horizons of the black hole hole, at the radial  locations $r_\pm = M \pm \sqrt{M^2 - a^2}$.  

The general form of the Kerr azimuthal 4-velocity is the following 
\beq
{{\rm d}\phi \over {\rm d} \tau} = U^\phi = {2M\ei a / r + \ji (1 - 2M/r) \over r^2 - 2Mr + a^2} ,
\eeq
which implies that the orbital integral takes the form
\beq
\sqrt{1 - \ei^2} \phi = \int {r^{3/2} (2M\ei a / r + \ji (1 - 2M/r)) \over (r - r_+)(r-r_-)(r - r_c) \sqrt{r_3 - r}} \, {\rm d} r .
\eeq
The solution of this integral is clearly more complicated than its Schwarzschild analogue. We define a  ``Kerr-type'' integral to be of the form 
\beq
I_{\cal K} = \int {x^{n/2} \over (x-A)(x-B)(x-C)\sqrt{D-x}} \, {\rm d}x .
\eeq
Kerr-type integrals can be transformed into Schwarzschild-type integrals by repeated use of the identity 
\beq
{1 \over (x-\alpha)(x-\beta)} = {1 \over \alpha-\beta} \left[ {1 \over x-\alpha} - {1 \over x - \beta}  \right] ,
\eeq
where a Schwarzschild-type integral has the following form 
\beq
I_{\cal S} = \int {x^{n/2} \over (x-\Gamma)\sqrt{\Delta-x}} \, {\rm d}x .
\eeq
The solutions of these Schwarzschild-type integrals can then be found by similar steps to those performed in section \ref{schwarz}. 

The following Kerr-integral solutions will be useful 
\begin{align}
I_1 &= \int {x^{3/2} \over (x-A)(x-B)(x-C)\sqrt{D-x}} \, {\rm d}x \nonumber \\
&=  {2A^{3/2} \over (A-B)(C-A)\sqrt{D-A}} \tanh^{-1} \left( \sqrt{D/x - 1 \over D/A - 1} \right) \nonumber \\
&- {2B^{3/2} \over (A-B)(C-B)\sqrt{D-B}} \tanh^{-1} \left(\sqrt{D/x - 1 \over D/B - 1} \right) \nonumber \\
&- {2C^{3/2} \over (C-A)(C-B)\sqrt{D-C}} \tanh^{-1} \left(\sqrt{D/x - 1 \over D/C - 1} \right) ,
\end{align}
and 
\begin{align}
I_2 &= \int {x^{1/2} \over (x-A)(x-B)(x-C)\sqrt{D-x}} \, {\rm d}x \nonumber \\
&= - {2A^{1/2} \over (A-B)(A-C)\sqrt{D-A}} \tanh^{-1}\left(\sqrt{D/x - 1 \over D/A - 1} \right)\nonumber \\
&+ {2B^{1/2} \over (A-B)(B-C)\sqrt{D-B}} \tanh^{-1}\left(\sqrt{D/x - 1 \over D/B - 1} \right)\nonumber \\
&- {2C^{1/2} \over (C-A)(C-B)\sqrt{D-C}} \tanh^{-1}\left(\sqrt{D/x - 1 \over D/C - 1} \right).
\end{align}
It is important here to note that these solutions assume $0 < A < B < C < D$, and $A, B, C < x < D$. For the case $C > D$ the identity $-i\tanh^{-1}(iz) = \tan^{-1}(z)$ allows the final term to be appropriately re-written. For $D=C$ some care must be taken, which is discussed further below.  The arguments of the $\tanh^{-1}$ functions should be inverted for $x < A, B, C$.

The general form of the Kerr orbital solution is therefore given by a sum of three functions,
\beq
\phi(r) = f_+(r) + f_-(r) + f_0(r) , 
\eeq
where the functions $f_\pm$ are defined as the functions which encapsulate the divergent $\phi$ behaviour at $r_\pm$, while $f_0$ will be qualitatively similar in character to the Schwarzschild solutions discussed above. For $r_c > r_{\rm ib}$, $f_\pm(r)$ have the following form 
\begin{multline}
f_\pm(r) = \pm \left[{2\ji r_\pm^{3/2} + 4Mr_\pm^{1/2} (a\ei - \ji) \over (r_+-r_-)(r_c - r_\pm)\sqrt{r_3 - r_\pm}} \right] \\ \times {1 \over \sqrt{1 - \ei^2}} \tanh^{-1} \left(\sqrt{r_3 / r - 1 \over r_3 / r_\pm - 1} \right).
\end{multline}
Note that for $a = 0$, where $r_+ = 2M, r_- = 0$, we find $f_\pm = 0$. For $r_c \leq r_{\rm ib}$ the form of $f_\pm(r)$ changes slightly, as will be discussed below. Note that for radii within the event horizons $r < r_\pm$ the arguments of the $\tanh^{-1}$ functions in $f_\pm$ should be inverted. 

\subsection{$r_c > r_{I}$ plunging orbits}
For $r_c > r_I$, the third root of the radial velocity polynomial $r_3$ is small $r_+ < r_3 < r_I < r_c$. In this limit the only solution for $f_0(r)$ is the following 
\begin{multline}
f_0(r) =  \left[{2\ji r_c^{3/2} + 4Mr_c^{1/2} (a\ei - \ji) \over (r_c-r_+)(r_c - r_-)\sqrt{r_c - r_3}} \right] \\ \times {1 \over \sqrt{1 - \ei^2}} \tan^{-1}\left(\sqrt{r_3 / r - 1 \over 1 - r_3 / r_c } \right),
\end{multline}
which is entirely analogous to the Schwarzschild plunge discussed above. Note that frame dragging and the poorly behaved $\phi$ coordinate on the Kerr event horizons mean that there is no well-defined ``rotated angle'' which this orbit traverses over its infall. However, there remain, in the formal limit $r_c \rightarrow \infty$,  orbital solutions of the Kerr equations which start outside of the event horizon at $r = 2M$, before subsequently reaching $r=0$, despite the particle having an arbitrarily large specific angular momentum. 

\subsection{$r_c = r_{I}$ spiral orbits}
For $r_c = r_I$, the third root of the radial velocity polynomial $r_3$ is equal to the ISCO $r_3 = r_c = r_I$. In this limit the  solution $f_0(r)$ is the following (first derived in \cite{MB})
\beq
f_0(r) =   \sqrt{6r_I\over M}\ \left( {2M(J-a\gamma) - r_IJ\over r^2_I -2Mr_I +a^2} \right) \sqrt{r\over r_I - r} .
\eeq

\subsection{$r_{\rm ib} < r_c < r_{I}$ homclinic orbits}
For $r_{\rm ib} < r_c < r_I$, the third root of the radial velocity polynomial $r_3$ is greater than the ISCO $r_3 > r_I > r_c$. In this limit there are two solutions for $f_0(r)$. The first, the homoclinic solution,  is the following (as first derived in \cite{LP1}) 
\begin{multline}
f_0(r) = - \left[{2\ji r_c^{3/2} + 4Mr_c^{1/2} (a\ei - \ji) \over (r_c-r_+)(r_c - r_-)\sqrt{r_3 - r_c}} \right] \\ \times {1 \over \sqrt{1 - \ei^2}} \tanh^{-1}\left(\sqrt{r_3 / r - 1 \over r_3 / r_c - 1} \right) .
\end{multline}

The second, a plunging orbit, has the following form:
\begin{multline}
f_0(r) = - \left[{2\ji r_c^{3/2} + 4Mr_c^{1/2} (a\ei - \ji) \over (r_c-r_+)(r_c - r_-)\sqrt{r_3 - r_c}} \right] \\ \times {1 \over \sqrt{1 - \ei^2}} \tanh^{-1}\left(\sqrt{ r_3 / r_c - 1 \over r_3 / r - 1} \right).
\end{multline}

\subsection{$r_c = r_{\rm ib}$ IBCO parabolic orbits}
For $r_c = r_{\rm ib}$, we have simultaneously $1 - \ei^2 = 0$ and $r_3 \rightarrow \infty$. The combination $\sqrt{(1 - \ei^2)(r_3 - r_c)}$ converges to $\sqrt{2M} (\ji - a)/r_{\rm ib}$, and the three functions have the following form 
\begin{multline}
f_\pm(r) = \pm \left[{2\ji r_{\rm ib} r_\pm^{3/2} + 4Mr_\pm^{1/2} r_{\rm ib} (a - \ji) \over \sqrt{2M}  (r_+-r_-)(r_{\rm ib} - r_\pm) (J - a)} \right]  \\ \times  \tanh^{-1} \left(\sqrt{r_\pm \over r }\right) .
\end{multline}
and
\begin{multline}
f_0(r) = - \left[{2\ji r_{\rm ib}^{5/2}  + 4M r_{\rm ib}^{3/2} (a - \ji) \over \sqrt{2M}  (r_{\rm ib}-r_+)(r_{\rm ib} - r_-) (J - a)} \right]  \\ \times  \tanh^{-1}\left(\sqrt{r_{\rm ib} \over r }\right) .
\end{multline}
for orbits which spiral outwards $r > r_{\rm ib}$. For inspiralling ($r<r_{\rm ib}$) orbits, we instead have 
\begin{multline}
f_0(r) = - \left[{2\ji r_{\rm ib}^{5/2} + 4M r_{\rm ib}^{3/2} (a - \ji) \over \sqrt{2M}  (r_{\rm ib}-r_+)(r_{\rm ib} - r_-) (J - a)} \right]  \\ \times  \tanh^{-1}\left(\sqrt{r \over r_{\rm ib} }\right) .
\end{multline}
For the inspiralling orbits, and for radii within the event horizons $r < r_\pm$, the arguments of the $\tanh^{-1}$ functions in $f_\pm$ should be inverted. 

\subsection{$r_{p} < r_c < r_{\rm ib}$ hyperbolic orbits}

 For $r_{p} < r_c < r_{\rm ib}$ we have $\gamma > 1, r_3 < 0$, and a test particle can escape to infinity with non-zero radial velocity.  The shape of the particle orbit is:

\begin{multline}
f_\pm(r) = \pm \left[{2\ji r_\pm^{3/2} + 4Mr_\pm^{1/2} (a\ei - \ji) \over (r_+-r_-)(r_c - r_\pm)\sqrt{|r_3| + r_\pm}} \right] \\ \times {1 \over \sqrt{ \ei^2 - 1 }} \tanh^{-1}\left(\sqrt{|r_3| / r + 1 \over |r_3| / r_\pm + 1}\right) ,
\end{multline}

and 

\begin{multline}
f_0(r) = - \left[{2\ji r_c^{3/2} + 4Mr_c^{1/2} (a\ei - \ji) \over (r_c-r_+)(r_c - r_-)\sqrt{|r_3| + r_c}} \right] \\ \times {1 \over \sqrt{ \ei^2 - 1 }} \tanh^{-1}\left(\sqrt{|r_3| / r + 1 \over |r_3| / r_c + 1}\right) ,
\end{multline}
for orbits which spiral outwards $r > r_{\rm c}$. The inspiralling orbits instead have 

\begin{multline}
f_0(r) = - \left[{2\ji r_c^{3/2} + 4Mr_c^{1/2} (a\ei - \ji) \over (r_c-r_+)(r_c - r_-)\sqrt{|r_3| + r_c}} \right] \\ \times {1 \over \sqrt{ \ei^2 - 1 }} \tanh^{-1}\left(\sqrt{ |r_3| / r_c + 1  \over |r_3| / r + 1}\right) .
\end{multline}

\section{Extremal Kerr Black holes, $a = \pm M$}
For $a = \pm M$, the two event horizons of the Kerr black hole coincide $r_\pm = M$, and so the partial fractions approach to solving the Kerr-type integrals must be modified. As far as the authors are aware, none of the pseudo-circular orbital solutions have been studied before in this extremal limit (except the ISCO inspiral \cite{MB}). 

\subsection{Extremal prograde spin $a = +M$}
For $a = +M$, circular orbits in the Boyer-Lindquist coordinate system are stable at all radii down to the event horizon $r_+ = M$. (In these coordinates, this is also the ISCO radius).  We therefore have $r_\pm < r_3 < r_c$ for all $r_c$, and there is only one valid pseudo-circular orbital solution, a plunging solution starting at $r_3$.   The solution no longer has two $f_\pm(r)$ solution, as the two event horizons coincide, but is of the form

\beq
\sqrt{1 - \ei^2} \phi = \int {r^{3/2} (2M^2\ei / r + \ji (1 - 2M/r)) \over (r - M)^2(r-r_c) \sqrt{r_3 - r}} \, {\rm d} r.
\eeq

\begin{figure}
\includegraphics[width=1.\linewidth]{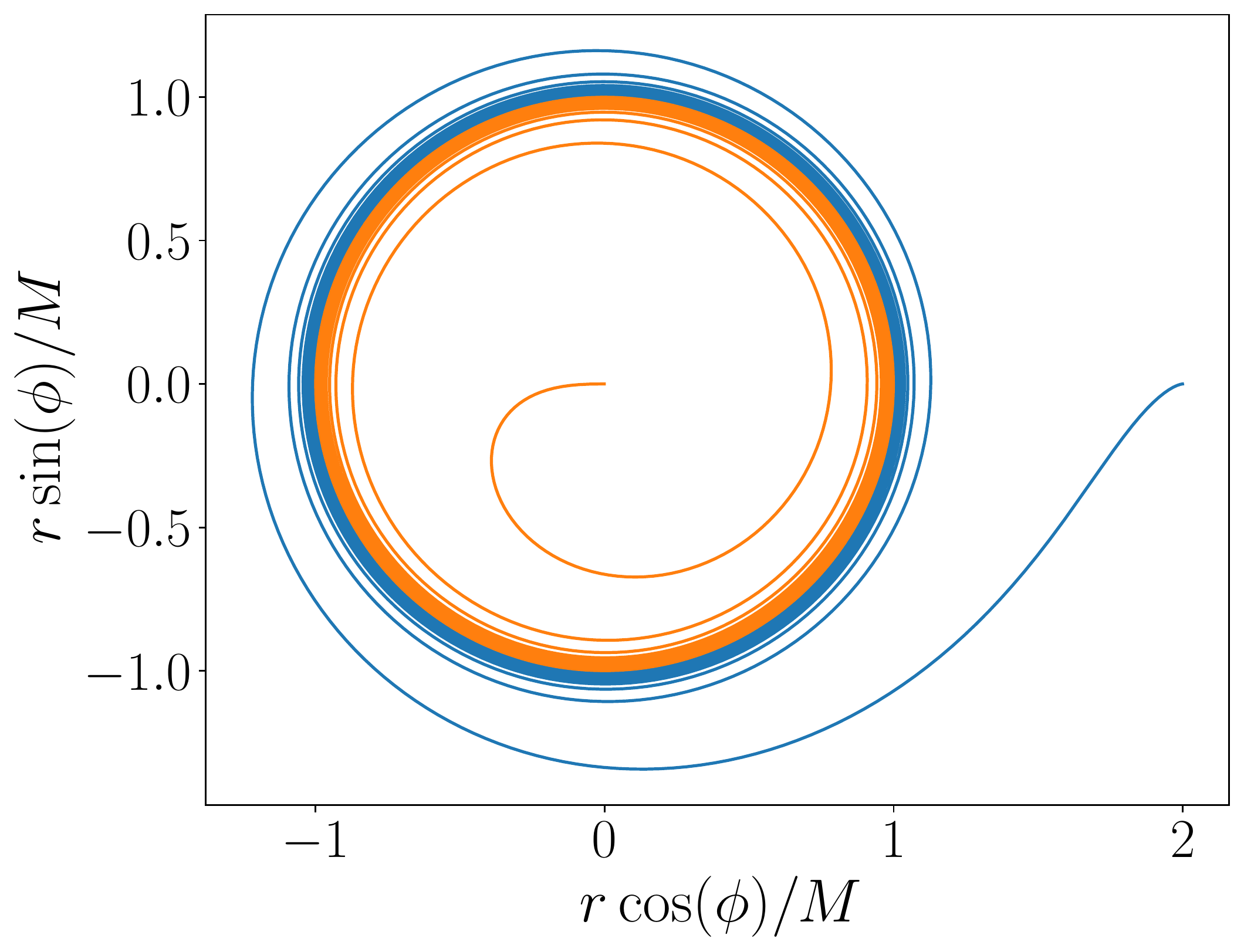}
\caption{An extremal prograde ($a = +M$) Kerr plunging  orbit, with $r_c = 10^6 M$. The colours denote the trajectory inside (orange) and outside (blue) the event horizons $r_\pm = M$. }
\label{extreme_plunge}
\end{figure}

The solution is again most simply written as a sum of three functions:

\beq
\phi(r) = f_0(r) + f_H(r) + \tilde f(r)
\eeq

where

\begin{multline}
f_0(r) =  \left[{2\ji r_c^{3/2} + 4Mr_c^{1/2} (M\ei - \ji) \over (r_c-M)^2\sqrt{r_c - r_3}} \right] \\ \times {1 \over \sqrt{1 - \ei^2}} \tan^{-1}\left(\sqrt{r_3 / r - 1 \over 1 - r_3 / r_c } \right),
\end{multline}
the analogue of the Schwarzschild plunge, 
and 
\beq
f_H(r) = - {M (\ji - 2M\ei ) \over (r_c - M)(r_3 - M) \sqrt{1 - \ei^2}}  {\sqrt{r(r_3 - r)} \over r-M} ,
\eeq
which encapsulates the divergent $\phi$ behaviour of the event horizon, and finally 
\begin{multline}
\tilde f(r) =  {M^{1/2} (1 - \ei^2)^{-1/2} \over  (r_3 - M)^{3/2}(r_c - M)^2} \tanh^{-1}\left(\sqrt{r_3 / r - 1 \over r_3 / M - 1}\right) \\ \times \bigg[ \ji (r_cr_3 - 2 r_c M - 3 r_3 M + 4M^2) \\ +  2M \ei (r_cr_3 + Mr_3 - 2M^2) \bigg]   ,
\end{multline}
for $r_3 > r > M$, and  with inverted $\tanh^{-1}$ argument
for $r < M$. In the formal limit $r_c \rightarrow \infty$,  $r_3 \to 2M$, and the above solution describes an orbital plunge from $2M$ to the singularity. An example trajectory, with circular orbit parameter $r_c = 10^6 M$, is displayed in Fig. \ref{extreme_plunge}.

\subsection{Extremal retrograde spin $a = -M$}
The extremal retrograde spin limit $a \to -M$ is a limit of great interest, as it highlights  some of the most interesting frame-dragging effects of the Kerr metric quite simply. 

For $a = -M$, the key orbital radii take the following values, $r_I = 9M$, $r_{\rm ib} = (3 + \sqrt{8})M/2$, $r_p = 4M$ and event horizons  $r_\pm = M$.   In this limit, note that all key radii are distinct and a full set of orbital solutions is available.

The orbital shape integral is of the form 
\beq\label{retint}
\sqrt{1 - \ei^2} \phi = \int {r^{3/2} (\ji (1 - 2M/r) - 2M^2\ei / r ) \over (r - M)^2(r-r_c) \sqrt{r_3 - r}} \, {\rm d} r .
\eeq
Once again the general solution comprises of three functions: 
\beq
\phi(r) = f_0(r) + f_H(r) + \tilde f (r) .
\eeq
The function
\beq
f_H(r) = - {M (\ji + 2M\ei ) \over (r_c - M)(r_3 - M) \sqrt{1 - \ei^2}}  {\sqrt{r(r_3 - r)} \over r-M} ,
\eeq
 for all $r_c \neq r_I, r_c > r_{\rm ib}$. It will, as will be discussed shortly,  be modified for $r_c < r_{\rm ib}$ and $r_c = r_I$.   The second function is equal to 
\begin{multline}
\tilde f(r) =  {M^{1/2} (1 - \ei^2)^{-1/2} \over  (r_3 - M)^{3/2}(r_c - M)^2} \tanh^{-1}\left( \sqrt{r_3 / r - 1 \over r_3 / M - 1} \right)\\ \times \bigg[ \ji (r_cr_3 - 2 r_c M - 3 r_3 M + 4M^2)\\ -  2M \ei (r_cr_3 + Mr_3 - 2M^2) \bigg]   ,
\end{multline}
which again will be modified slightly for intra-IBCO circular orbits, and for $r_c = r_I$. 
Note that for the special ISCO location, where $r_c = r_3 = r_I = 9M$ (with $\ji = 22\sqrt{3}M/9$, $\ei = 5\sqrt{3}/9$) the function $\tilde f (r) = 0$ as the final term in the square brackets vanishes. We now discuss the eight distinct orbital solutions of the extremal retrograde Kerr metric. 

\subsubsection{$r_c > r_{I}$ plunging orbits}
The plunging orbital solution for extremal $a = -M$ is  given  by: 
\begin{multline}
f_0(r) =  \left[{2\ji r_c^{3/2} - 4Mr_c^{1/2} (M\ei + \ji) \over (r_c-M)^2\sqrt{r_c - r_3}} \right] \\ \times {1 \over \sqrt{1 - \ei^2}} \tan^{-1}\left( \sqrt{r_3 / r - 1 \over 1-r_3 / r_c} \right) .
\end{multline}

\subsubsection{$r_c = r_{I}$ spiral orbits}
For $r_c = r_I$, the third root of the radial velocity polynomial $r_3$ is equal to the ISCO $r_3 = r_c = r_I$. In this limit the solution for $\phi(r)$ is  \cite{MB}:
\beq
\phi(r) =   {2\sqrt{2} \over 3 M^{3/2} } {r^{3/2} \over (1 - r/M) \sqrt{1 - r/9M}} ,
\eeq
which is most easily found by returning to the governing integral (\ref{retint}) and simplifying in this limit.

\subsubsection{$r_{\rm ib} < r_c < r_{I}$ homclinic orbits}
The homoclinic orbital solution for extremal $a = -M$ is found to be:

\begin{multline}
f_0(r) =  \left[{2\ji r_c^{3/2} - 4Mr_c^{1/2} (M\ei + \ji) \over (r_c-M)^2\sqrt{r_3 - r_c}} \right] \\ \times {1 \over \sqrt{1 - \ei^2}} \tanh^{-1}\left( \sqrt{r_3 / r - 1 \over r_3 / r_c - 1} \right),
\end{multline}
as is the corresponding second orbit, a plunging orbit:
\begin{multline}
f_0(r) =  \left[{2\ji r_c^{3/2} - 4Mr_c^{1/2} (M\ei + \ji) \over (r_c-M)^2\sqrt{r_3 - r_c}} \right] \\ \times {1 \over \sqrt{1 - \ei^2}} \tanh^{-1}\left( \sqrt{r_3 / r_c - 1 \over r_3 / r - 1} \right),
\end{multline}

\subsubsection{$r_c = r_{\rm ib}$ IBCO parabolic orbits}
For $r_c = r_{\rm ib}$, we have simultaneously $1 - \ei^2 = 0$ and $r_3 \rightarrow \infty$. The combination $\sqrt{(1 - \ei^2)(r_3 - r_c)}$ converges to $\sqrt{2M} (\ji + M)/r_{\rm ib}$, and the three functions have the following form (presented here for the inspiralling case) 
\beq
\phi(r) = \alpha \tanh^{-1}\sqrt{M \over r} + \beta \tanh^{-1}\sqrt{r\over r_{\rm ib}} + \gamma {r^{1/2} \over r-M} .
\eeq
The argument $\alpha \tanh^{-1}$ should of course be inverted within the event horizon $r<M$. 
The three constants are  
\begin{align}
\alpha &= - { r_{\rm ib} J (M - 3r_{\rm ib}) + 2 r_{\rm ib} (r_{\rm ib} + M)(J + M) \over \sqrt{2}(r_{\rm ib}-M)^2(J + M) }, \nonumber \\
\beta &= { 2\sqrt{2} r_{\rm ib}^{3/2} M(J+M) -\sqrt{2} r_{\rm ib}^{5/2}J \over \sqrt{M}(r_{\rm ib}-M)^2(J + M) }, \nonumber \\
\gamma &= {3 r_{\rm ib} J M^{1/2} + 2 r_{\rm ib} M^{3/2} \over \sqrt{2}(r_{\rm ib}-M)(J + M) } .
\end{align}
The argument of $\beta \tanh^{-1}$ should be inverted for outspiralling orbits, where $r > r_{\rm ib}$.
\subsubsection{$r_{p} < r_c < r_{\rm ib}$ hyperbolic orbits}
For $r_{p} < r_c < r_{\rm ib}$ we have $\gamma > 1, r_3 < 0$, and a test particle can escape to infinity with non-zero radial velocity.  Each of the three orbital functions must be modified slightly in this limit. They are
\beq
f_H(r) =  {M (\ji + 2M\ei ) \over (r_c - M)(|r_3| + M) \sqrt{ \ei^2 - 1 }}  {\sqrt{r(|r_3| + r)} \over r-M} ,
\eeq
and
\begin{multline}
\tilde f(r) =  {- M^{1/2} ( \ei^2 - 1 )^{-1/2} \over  (|r_3| + M)^{3/2}(r_c - M)^2 } \tanh^{-1}\left( \sqrt{|r_3| / r + 1 \over |r_3| / M + 1} \right)\\ \times \bigg[ \ji (r_cr_3 - 2 r_c M - 3 r_3 M + 4M^2) \\ -  2M \ei (r_cr_3 + Mr_3 - 2M^2) \bigg]   ,
\end{multline}
and finally 
\begin{multline}
f_0(r) =  \left[{2\ji r_c^{3/2} - 4Mr_c^{1/2} (M\ei + \ji) \over (r_c-M)^2\sqrt{r_c + |r_3|}} \right] \\ \times {1 \over \sqrt{ \ei^2 - 1 }} \tanh^{-1}\left(\sqrt{|r_3| / r + 1 \over |r_3| / r_c + 1}\right) ,
\end{multline}
for the outspiralling orbit $r > r_c$. The corresponding inspiralling orbit is instead described by a function of the same form, but with its $\tanh^{-1}$ argument inverted.
This concludes the set of massive particle pseudo-circular orbits. 
 
\section{ Radial momentum and orbits of photons }
The case of massless particles, which for simplicity we refer to generically as ``photons'', proceeds along similar lines. The trajectories of photon orbits are well studied in the literature, with the Schwarzschild solutions first presented in \cite{Darwin}, and a general discussion of Kerr photon orbits is presented in \cite{Chand}. The case of extremal Kerr orbits, which are not simple limits of the more general $|a| \neq M$ Kerr solutions, do not appear to have been derived previously, and are presented here.     For completeness we also present the Schwarzschild solutions of \cite{Darwin}. 

We work with the 4-momenta $p^\mu ={\rm d}x^\mu/{\rm d}\sigma$, where $\sigma$ is a suitable affine parameter.  The equations for a circular photon orbit are now
\beq\label{photos}
g_{\mu\nu} p^\mu p^\nu = 0 ,  \quad p^\mu p^\nu \dd_r g_{\mu\nu} = 0 ,
\eeq
with only $p^0$ and $p^\phi$ present.   The conserved energy and angular momentum correspond to (minus) $p_0$ and $p_\phi$.   The solution of this homogeneous system requires the determinant of the coefficients to vanish, which is the same as solving the following 
\beq
\D^2=1 -3M/r_p +2aM^{1/2}/r_p^{3/2} = 0 .
\eeq
Solving this with the standard cubic formula gives the radius of a circular photon orbit:
\beq
r_p = 4M\cos^{2}\left[{1\over3}\cos^{-1}\left(-{a \over M}\right)\right]
\eeq
which varies from $4M$ to $M$ for $-1\le a/M \le 1$.    With $\D=0$, we may now solve for the (conserved) ratio $p_\phi/p_0$ from either one of the equations in (\ref{photos}), the first being more convenient.     The end result is:
\beq
p_\phi = (a - 3\sqrt{M r_p}) p_0.
\eeq
(To arrive at this apparently simple result is non-trivial because it involves very selectively substituting for $a$ in terms of $r_p$ using the original ${\cal D}=0$ equation.    See \cite{HEL} for an alternative derivation.)  

Next, we retain the conserved circular $p_0$ (which is arbitrary) and $p_\phi$ (from above) values, while also allowing for a $p^r$ component to be present.  Our governing equation ($p^\mu p_\mu =0$) is then:
\begin{multline}\label{photonfull}
r^2 \left({p^r \over p_0}\right)^2 - {p_\phi \over p_0} \left( {2Ma \over r} - \left(1 - {2M \over r}\right) {p_\phi \over p_0} \right) \\ - \left(\left(r^2 + a^2 + {2Ma^2 \over r}\right) + {2Ma \over  r} {p_\phi \over p_0}\right) = 0.
\end{multline}
Proceeding as before, this unwieldy equation may be factored in terms of the roots $r_p$ and $r_3$ by defining an effective potential.  Matters simplify dramatically: 
\beq
\left({p^r \over p_0}\right)^2 = \left(1 - {r_p \over r} \right)^2 \left(1 - {r_3 \over r}\right) .
\eeq
(The normalisation follows from equation \ref{photonfull} by taking the limit $r\rightarrow \infty$.)  
Taking the limit $\lim_{r\to0} r^3 (p^r/p_0)^2$ gives, from the above equation, $- r_p^2 r_3$, and from the governing equation \ref{photonfull}
\beq
-r_p^2 r_3 = 2M \left({p_\phi \over p_0} + a\right)^2 = 2M (3\sqrt{Mr_p} - 2a)^2 = 2 r_p^3
\eeq 
where we used (from ${\cal D} = 0$)
\beq
2a = 3\sqrt{Mr_p} - r_p^{3/2} M^{-1/2} ,
\eeq
to arrive at the final equality. Thus, the remarkable solution from the null geodesic equation that emerges is
\beq
(p^r)^2 = (1+2r_p/r)(1-r_p/r)^2 (p_0)^2.
\eeq
In common with the case of massive particles inspiralling from the ISCO \cite{MB}, this depends on $a$ only through the photon orbit radius $r_p$.  This orbital equation is of the same form as the massive particle hyperbolic orbit. 

To determine the photon orbits spiralling from the photon radius, we require $p^\phi$, which is readily evaluated from $p_\phi$ and $p_0$:
\begin{align}
p^\phi  & =  {p_0\over \Delta}\left( a -{4Ma\over r} - 3({Mr_p})^{1/2} + {6M^{3/2}r^{1/2}_p \over r}\right), \nonumber\\
& = {p_0\over \Delta}\left( a - 3(Mr_p)^{1/2} + {2M^{1/2}r^{3/2}_p \over r}\right) , 
\end{align}
where in the final line we have used ${\cal D} =0$ to substitute for $4Ma/r$, and $\Delta = (r - r_+)(r-r_-)$. 
\subsection{Schwarzschild photon spirals} 
 In the Schwarzschild $a= 0$ limit, $p^\phi = -3\sqrt{3}Mp_0/r^2$.    Then with $r_p=3M$, $x=r/r_p$,
\beq
{{\rm d} \phi\over {\rm d}x} = r_p{{\rm d} \phi \over {\rm d}r} = r_p{p^\phi\over p^r} = - {1\over x-1}\sqrt{3\over x(x+2)}
\eeq
Then
\begin{align}
\phi & = - \int {{\rm d}x\over x-1}\sqrt{3\over x(x+2)}\nonumber\\
& = 2\tanh^{-1}\sqrt{3x\over x+2}, \quad (x<1), \nonumber\\
& =  2\tanh^{-1}\sqrt{x+2\over 3x}, \quad (x>1). 
\end{align}
Solving for $r$:
\begin{align}
r &= {6M\over 3\coth^2(\phi/2) - 1}\quad {\rm (inspiral)}, \nonumber\\
& = {6M\over 3\tanh^2(\phi/2) - 1}\quad {\rm (outspiral)}.
\end{align}
These solutions formed  part of Darwin's 1959 analysis \cite{Darwin}. 
Note that as in the case of the massive particle hyperbolic orbits, photons following this outspiralling orbit reach infinity at a finite angle 
\beq
\phi_\infty = - 2 \tanh^{-1}\left(\sqrt{1\over 3}\right) .
\eeq
\subsection{General Kerr solution $-M < a < M$}
The general Kerr photon integral is of the following form 
\beq
\phi(r) = \int {2M^{1/2} r_p^{3/2} r^{1/2} + (a - 3\sqrt{Mr_p})r^{3/2} \over (r-r_+)(r-r_-)(r-r_p)\sqrt{r + 2r_p}} \, {\rm d} r,
\eeq
whose solution can be found via a partial fraction expansion. Omitting the lengthy but straightforward details, we find 
\begin{multline}
\phi(r) = C_0 \tanh^{-1}\sqrt{3r \over 2r_p + r} + C_+ \tanh^{-1}\sqrt{1+2r_p/r \over 1 + 2r_p / r_+} \\
+ C_- \tanh^{-1}\sqrt{1+2r_p/r \over 1 + 2r_p / r_-} 
\end{multline}
For an inspiralling $r_\pm < r < r_p$ solution (the arguments of the final two $\tanh^{-1}$ functions are inverted for $r < r_\pm$). While the outspiraling solution $r > r_p$ is 
\begin{multline}
\phi(r) = C_0 \tanh^{-1}\sqrt{2r_p + r \over 3r} + C_+ \tanh^{-1}\sqrt{1+2r_p/r \over 1 + 2r_p / r_+} \\
+ C_- \tanh^{-1}\sqrt{1+2r_p/r \over 1 + 2r_p / r_-} .
\end{multline}
The coefficients $C_0, C_\pm$ of these solutions are the following
\beq
C_0 = {-4M^{1/2} r_p^{3/2} - 2r_p(a - 3\sqrt{Mr_p}) \over \sqrt{3} (r_p-r_+)(r_p-r_-)}
\eeq
and
\beq
C_\pm = \pm {4M^{1/2} r_p^{3/2}r_\pm^{1/2} + 2r_\pm^{3/2}(a - 3\sqrt{Mr_p}) \over  (r_+-r_-)(r_p-r_\pm)\sqrt{r_\pm + 2r_p}}
\eeq
Note that for $a = 0, r_p = 3M$, $r_+ = 2M$, $r_- = 0$, and therefore  $C_0 = 2$, $C_\pm = 0$.

\subsection{Extremal Kerr solutions $ a = \pm M$} 
For an extremal spin $a = \pm M$, the two event horizons of the Kerr black hole coincide $r_\pm = M$, and the partial fractions approach used in the previous section must be revisited. These solutions appear to have not been previously discussed in the literature, and display some remarkable properties. 

\subsubsection{Extremal retrograde spin $ a = -M$}
For the extremal retrograde spin of $a = -M$, the photon radius is at $r_p = 4M$, and the photons orbit has the following shape
\beq
\phi = \int {16 M^2 r^{1/2} - 7Mr^{3/2} \over (r - M)^2 (r - 4M) \sqrt{r + 8M}} \, {\rm d}r ,
\eeq
which has two possible solutions. For $r < 4M$, we have
\begin{multline}
\phi(r) = {1 \over 9} \Bigg[ {\sqrt{9r(8M + r)} \over r - M}  + 8\sqrt{3} \tanh^{-1}\sqrt{{r\over 4M}}  \\ + 4\sqrt{3} \ln\left({4  + \sqrt{r/M} + \sqrt{24 + 8r/M} \over 4 - \sqrt{r/M} + \sqrt{24 + 8r/M}}\right) \Bigg]
\end{multline}

while for $r > 4M$ we have

\begin{multline}
\phi(r) = {1 \over 9} \Bigg[ {\sqrt{9r(8M + r)} \over r - M}  + 8\sqrt{3} \tanh^{-1}\sqrt{{4M\over r}}  \\ + 4\sqrt{3} \ln\left({4  + \sqrt{r/M} + \sqrt{24 + 8r/M} \over 4 - \sqrt{r/M} + \sqrt{24 + 8r/M}}\right) \Bigg] . 
\end{multline}

\begin{figure}
\includegraphics[width=1.15\linewidth]{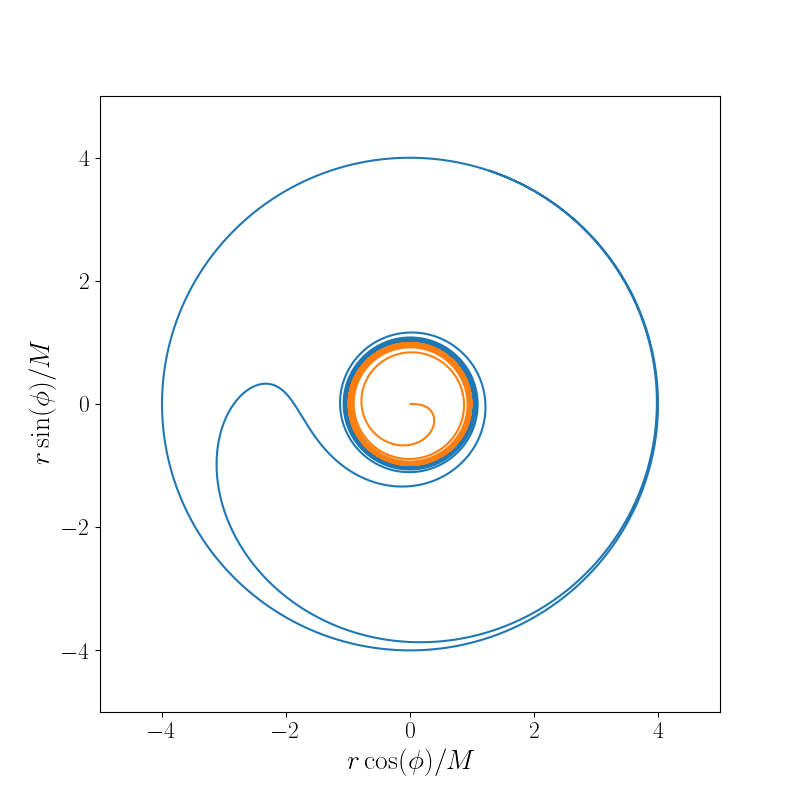}
\caption{The inspiralling photon trajectory for the extremal retrograde $a = -M$ Kerr metric. Pronounced frame dragging is clearly visible, with the trajectory changing direction just outside of the ergo-region $r_E = 2M$. }
\label{extreme_photon_ret}
\end{figure}

The inspiralling orbit displays pronounced frame dragging, and is displayed in Fig. \ref{extreme_photon_ret}. 

\subsubsection{Extremal prograde spin $ a = +M$}

\begin{figure}
\includegraphics[width=1.15\linewidth]{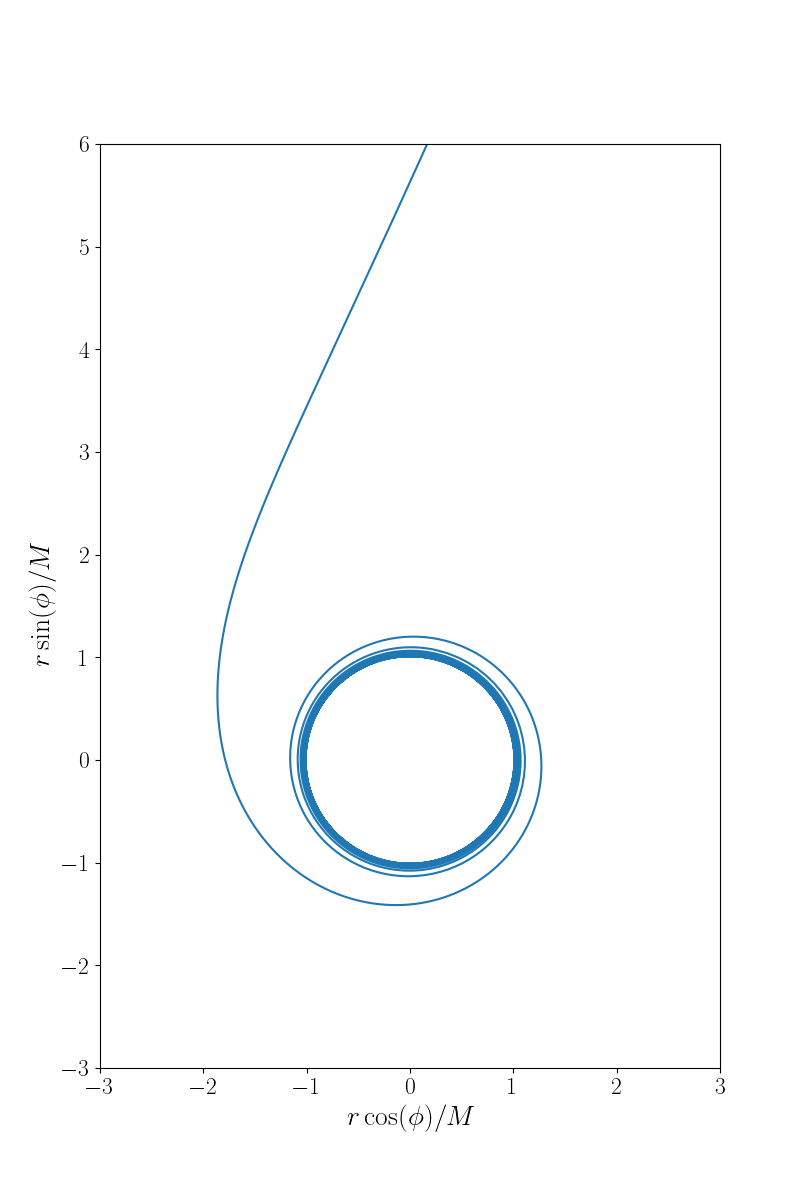}
\caption{The outspiralling photon trajectory for the extremal prograde $a = +M$ Kerr metric. This orbit starts formally at Boyer-Lindquist radial coordinate $r=M$, before escaping to infinity. }
\label{extreme_photon}
\end{figure}

For $a = +M$ the photon radius $r_p$ coincides with the two event horizon radii $r_\pm$, all occurring at the Boyer-Lindquist co-ordinate $r=M$. For a photon, there is in fact an orbital solution starting at the Boyer-Lindquist event horizon $r_+ = M$, which then escapes to infinity. With $r_p = r_\pm = a = M$, this orbital solution satisfies the following integral 
\beq
\phi = 2M \int{M r^{1/2} - r^{3/2} \over (r-M)^3 \sqrt{r+2M}} \, {\rm d}r ,
\eeq
which has the following solution 
\begin{multline}
\phi(r) = {2 \over 3 \sqrt{3}} \Bigg[ {\sqrt{3r(2M + r)} \over r - M}  + 2 \tanh^{-1}\sqrt{{M\over r}}  \\ +  \ln\left({2 + \sqrt{r/M} + \sqrt{6 + 3r/M} \over 2 - \sqrt{r/M} + \sqrt{6 + 3r/M}}\right) \Bigg] .
\end{multline}
This orbital solution is displayed in Fig. \ref{extreme_photon}. 

\section{Conclusions}
The pseudo-circular orbital solutions of the Schwarzschild, Kerr, and extremal Kerr, metrics, which are characterised by constants of motion that also describe a circular orbit, have been explicitly derived. There are eight such orbital classes, five of which plunge and terminate at the origin, while three either remain bound (homoclinic orbits), or escape to infinity with zero (parabolic) or finite (hyperbolic) radial velocities. Each orbital solution may be written entirely in terms of elementary functions, a result of the remarkable symmetry imposed on the orbits by the circular constants of motion. The in- or out-spiralling behaviour of photon trajectories starting at the unstable photon circular radius have also been discussed, and have qualitatively similar properties to the hyperbolic class of massive particle orbits.  

The systematic derivation of all eight orbital classes in this paper completes a line of analysis which dates back to Darwin's 1959 study of Schwarzschild orbital motion  \cite{Darwin}.  Although the origins of this field are more than half a century old, these explicit solutions have important modern applications, e.g. as contextual benchmarks for numerical accretion \cite{SK} or photon ray tracing problems \cite{PJ}, for the computation of gravitational waveforms arising from extreme mass ratio inspirals using the Teukolsky formalism \cite{T}, or in extending existing theories of black hole accretion \cite{Rey}.

\section*{Acknowledgments }
This work was supported by a Leverhulme Trust International Professorship grant [number LIP-202-014]. For the purpose of Open Access, AM has applied a CC BY public copyright licence to any Author Accepted Manuscript version arising from this submission. This work is partially supported by the Hintze Family Charitable Trust and STFC grant ST/S000488/1.

\label{lastpage}

\newpage 
\begin{references}
\bibitem{Bard} Bardeen J.~M., Press W.~H., Teukolsky S.~A., 1972, ApJ, 178, 347
\bibitem{Chand} Chandrasekhar, S.\ 1983, Mathematical Theory of Black Holes (Oxford: Clarendon Press)
\bibitem{CH} Collins N. A.,  Hughes S. A., 2004, Phys. Rev. D69, 124022
\bibitem{rev1} Comp{\`e}re G., Liu Y., Long J., 2022, PhRvD, 105, 024075 
\bibitem{Darwin} Darwin C., 1959, RSPSA, 249, 180
\bibitem{D} Drasco S., 2006, Class. Quant. Grav. 23, S769 
\bibitem{DH1} Drasco S., Hughes  S. A., 2004, Phys. Rev. D 69, 044015
\bibitem{DH2}  Drasco E. F. S.,  Hughes S. A., 2005, Class. Quant. Grav., 22, 801
\bibitem{DH3} Drasco S., Hughes S., 2006, Phys. Rev. D 73, 024027
\bibitem{FH} Fujita R., Hikida W., 2009, CQGra, 26, 135002
\bibitem{GHK} Glampedakis K., Hughes  S. A.,  Kennefick D., 2002, Phys. Rev. D 66, 064005.
\bibitem{G}   Glampedakis K., 2005,  Class. Quant. Grav. 22, S605 
\bibitem{HEL} Hobson, M.\ P., Efstahiou, G.,  Lazenby, A.\ N.\ 2006,  General Relativity: An Introduction for Physicists (Cambridge: Cambridge University Press)
\bibitem{LP1} Levin J., Perez-Giz G., 2009, PhRvD, 79, 124013
\bibitem{LP2} Levin J., Perez-Giz G., 2008, PhRvD, 77, 103005 
\bibitem{LH}  Lang R. N.,  Hughes S. A., 2006, Phys. Rev. D 74, 122001
\bibitem{MB} Mummery, A., Balbus, S.A. 2022, PRL, 129, 161101
\bibitem{NT} Novikov, I.\ D., Thorne, K.\ S.\ 1973, in {\sl Black Holes,} ed C.\ DeWitt \& B.\ DeWitt (New York: Gordon \& Breach)
\bibitem{PJ}{Psaltis D., Johannsen T., 2012, ApJ, 745, 1}
\bibitem{PT} Page D.~N., Thorne K.~S., 1974, ApJ, 191, 499
\bibitem{Rey} Reynolds, C.\ S.\, Begelman, M.\ C.\ 1997, ApJ, 487, 135
\bibitem{SK} Schnittman, J.\ D., Krolik, J.\ H., Noble, S. C. 2016, ApJ, 819, 48
\bibitem{T} Teukolsky S A, 1973, Astrophys. J. 185 635
\end{references}
\end{document}